\documentclass{article}
\usepackage[dvips]{graphicx}
\begin{document}

\title{Scaling Cosmology}
\author{Winfried Zimdahl\footnote{Electronic address:
winfried.zimdahl@uni-konstanz.de}\\
Fachbereich Physik, Universit\"at Konstanz\\ PF M678, 
D-78457 Konstanz, Germany\\
and\\
Diego Pav\'{o}n\footnote{Electronic address:  
diego@ulises.uab.es}\\
Departamento de F\'{\i}sica, 
Universidad Aut\'{o}noma de Barcelona\\
08193 Bellaterra (Barcelona), Spain}
\date{\today}
\maketitle

\begin{abstract}
We show that with the help of a suitable coupling between dark energy
and cold dark matter it is possible to reproduce any scaling solution
$\rho _{X}\propto \rho _{M}a ^{\xi }$, where $\rho _{X}$  
and $\rho_{M}$ are the densities of dark energy and dark matter, respectively. 
We demonstrate how the case $\xi = 1$ alleviates the coincidence problem. 
Future observations of supernovae at high redshift as well as  
quasar pairs which are planned to discriminate between different 
cosmological models will also provide direct constraints on the 
coupling between dark matter and dark energy.
\end{abstract}

\section{Introduction}
As is widely known, current observational evidence heavily favors an
accelerating and spatially flat Friedmann-Lema\^{\i}tre-Robertson-Walker universe
(for a pedagogical short update see \cite{update}). Since normal
matter fulfils the strong energy condition and cannot drive cosmic
acceleration, recourse is often made either to a small cosmological
constant ($\Lambda$CDM model) or to an almost evenly distributed source
of energy called ``dark energy" or ``quintessence" with equation of
state $p_{X} = w_{X} \rho_{X}$ where $-1\leq w_{X} < 0 $, such that it
makes the pressure negative enough to render the deceleration
parameter negative (see e.g. \cite{enough}). (Obviously, the quantity
$w_{X}$ depends on the particular form assumed by the potential of
the self-interacting quintessence scalar field). Since cold dark
matter (i.e., dust) and quintessence decay with the expansion at
different rates the question arises: ``why the ratio between CDM and
quintessence energies should be of the same order today?". In other
words, ``where the relationship $(\rho_{M}/\rho_{X})_{0} = {\cal
O}(1)$ comes from?" This is in essence the ``coincidence problem"
\cite{coincidence}. As usual, the zero subindex means present time.

Certain kind of models invoke that both components (dark matter and
dark energy) may not  be separately conserved due to some (unknown)
coupling between each other. This proposal has been explored
\cite{amendola}, \cite{interacting} and looks promising as a 
suitable mutual interaction can make both components redshift 
coherently. However,  because of problems of their own (as the 
inability to recover the dust era when going back in time) 
neither of these proposals, as they stand, can be regarded 
as the final answer.

As suggested by Dalal {\it et al.} it seems rather more advisable to
use the scanty observational information we possess to constrain the
quintessence field from a minimum of theoretical input than trying to
get a detailed fit to these data from any specific potential
\cite{dalal}. These authors introduced a generalized class of dark
energy models characterized by a non-canonical scaling of the ratio 
of the densities of dark matter and dark energy with the scale 
factor of the Robertson-Walker metric. They suggest a 
phenomenological form
\\
\begin{equation}
\frac{\rho_M}{\rho_X} \propto a^{-\xi}
\label{1}
\end{equation}
\\
for the ratio of the dark matter density $\rho_M$ to the density
$\rho_X$ of the dark energy, where the scaling parameter $\xi$ is
regarded as a new variable.  For an equation of state $p_X = -
\rho_X$ of the dark energy component a value $\xi =3$ amounts to the
$\Lambda$CDM model.  A value $\xi =0$ represents a stationary ratio
$\rho_M /\rho_X = {\rm const} $.

If the cosmological dynamics admits a stable, stationary solution
$\rho_M /\rho_X = {\rm const} $, corresponding to $\xi =0$ and the
present universe is already close to this state, there will be no
coincidence problem.  Consequently, according to \cite{dalal}, the
deviation of the parameter $\xi$ from $\xi =0$ quantifies the
severity of the problem. But it is not only the stationary solution
which deserves interest. Any solution which deviates from $\xi =
-3w_X$ represents a testable,
non-standard cosmological model and any solution with a scaling
parameter $\xi < 3$ will make the coincidence problem less severe.
It is therefore desirable to have a physical mechanism that could
give rise  to such kind of deviations from the standard dynamics.

The purpose of this paper is to show that a departure from the
standard $\xi = -3w_X$ case can be obtained if cold dark matter and 
quintessential dark energy are no longer assumed to be separately conserved.  
More precisely, we shall demonstrate that a suitable interaction between
dark matter and dark energy is able to produce any desired scaling.
The specific parameter choice $w_X =-3$ and $\xi=1$ is used to
establish an exactly solvable toy model for a non-standard
cosmological dynamics.  Upcoming observations which will constrain
cosmological models in a $\xi - w_X$ plane, as discusseed in
\cite{dalal}, will also put limits on such type of interactions.

\section{Scaling solutions}
We investigate a two--component system of cold dark matter (subindex $M$) and
dark energy (subindex $X$) where 
\\
\begin{equation}
\rho = \rho_{M} + \rho_{X}\ 
\qquad  {\rm and}\qquad
p = p_{M} + p_{X}\
\label{2}
\end{equation}
\\
are the total energy density and the total pressure, respectively.
The components are assumed to possess the equations of state 
\\
\begin{equation}
p _{M} \ll \rho _{M } \quad {\rm and} \quad 
p _{X} = w _{X}\rho _{X}\ .
\label{3}
\end{equation}
\\
We admit interactions between both components according to 
\\
\begin{equation}
\dot{\rho}_{M} + 3H \rho_{M} = Q
\label{4}
\end{equation}
\\
and 
\begin{equation}
\dot{\rho}_{X} 
+ 3H \left(1+w _{X}\right)\rho _{X} = -Q \ ,
\label{5}
\end{equation}
\\
where the coupling term $Q$ is to be determined below.
It is convenient to introduce the quantities 
$\Pi_{M}$ and $\Pi_{X}$ by 
\\
\begin{equation}
Q \equiv - 3H \Pi_{M} 
\equiv 3H \Pi_{X}\ ,
\label{6}
\end{equation}
\\
with the help of which we can write ($A$ = M, X)
\begin{equation}
\dot{\rho}_{A} + 3H \left(\rho_{A}
+ P _{A}\right) = 0
\ ,\quad 
P _{A}= p_{A} + \Pi_{A}
\ .
\label{7}
\end{equation}
The coupling is then included via 
$\Pi_{M}=-\Pi_{X}$. 

To derive a specific expression for the interaction term let us consider the 
time evolution of the ratio $\rho _{M}/\rho _{X}$,  
\\
\begin{equation}
\left(\frac{\rho _{M}}{\rho _{X}} \right)^{\displaystyle \cdot}
= \frac{\rho _{M}}{\rho _{X}}
\left[\frac{\dot{\rho }_{M}}{\rho _{M}} 
- \frac{\dot{\rho }_{X}}{\rho _{X}}\right]\ .
\label{8}
\end{equation}
\\
>From Eqs. (\ref{4})-(\ref{6})   we obtain
\begin{equation}
\left(\frac{\rho _{M}}{\rho _{X}} \right)^{\displaystyle \cdot}
= 3H \frac{\rho _{M}}{\rho _{X}}
\left[w _{X} 
- \frac{\rho }{\rho _{M}\rho _{X}}
\Pi _{M}\right]\ .
\label{9}
\end{equation}
We look for solutions with the scaling behavior 
\begin{equation}
\frac{\rho _{M}}{\rho _{X}} 
= r\left(\frac{a _{0}}{a} \right)^{\xi }\ .
\label{10}
\end{equation}
\\
Here, $r$ denotes the ratio of both components at the present time , i.e.,
at $a =a _{0}$, and the parameter $\xi$ is a constant. 
Inserting (\ref{10}) into (\ref{9})  and solving for $\Pi _{M}$ we find 
\\ 
\begin{equation}
\Pi _{M} = - \Pi _{X}
= \left[\frac{\xi }{3} + w _{X}\right]
\frac{a ^{\xi }}{a ^{\xi }+ra _{0}^{\xi }}  
\rho _{M} 
= \frac{\frac{\xi }{3} + w _{X}}
{1 + r \left(1+z \right)^{\xi }} \rho _{M}
\ ,
\label{11}
\end{equation}
\\
where $1+z \equiv  a _{0}/a $.  
This generalizes previous investigations for the case $\xi =0$ 
\cite{interacting}.  There is a transfer of energy from the scalar field to
the matter, i.e., $Q >0$, for $w _{X} + (\xi /3) < 0 $.
 
$\Pi _{M}$ and $\Pi _{X}$  are the effective pressures, equivalent to
those interaction between both components given by the quantity $Q$
in (\ref{6}),  which guarantee a scaling solution (\ref{10}).  We
arrive at the conclusion that {\it by a suitable choice of the
interaction between both components we may produce any desired
scaling behavior of the energy densities.}\\ The uncoupled case
corresponding to $\Pi _{M} = 0$  is given by
\\
\begin{equation}
\frac{\xi }{3} + w _{X} = 0 \ .
\label{12}
\end{equation}
\\
The $\Lambda $CDM model is recovered as the special case with 
$w _{X}=-1$ and $\xi =3$.    
The interacting models are parametrized by deviations from $\xi =-3w_X$, 
equivalent to deviations from $\xi=3$ for $w_X =-1$. 
>From (\ref{11}) we find 
\\
\begin{equation}
\frac{\Pi _{M}}{\rho _{M}} = \frac{\frac{\xi }{3} + w _{X}}
{1 + r \left(1+z \right)^{\xi }} \ ,
\label{13}
\end{equation}
\\
and
\\ 
\begin{equation}
-\frac{\Pi _{X}}{\rho _{X}} = r\left(\frac{\xi }{3} + w _{X}\right)
\frac{\left(1+z \right)^{\xi }}
{1 + r \left(1+z \right)^{\xi }} \ .
\label{14}
\end{equation}
\\
In the following we shall assume $\xi \neq -3w_X$ and $\xi > 0$, i.e., we
consider departures from the standard case of separately conserved
quantities.  For $1 \ll z $, i.e., when $\rho_X \ll \rho_M$ (according to (\ref{10})), we have
\\
\begin{equation}
\frac{|\Pi _{M}|}{\rho _{M}} \ll 1
\qquad\qquad\qquad\qquad
(1 \ll z), 
\label{15}
\end{equation}
\\
and 
\begin{equation}
-\frac{\Pi _{X}}{\rho _{X}} = \frac{\xi }{3} + w _{X}
\qquad\qquad
(1 \ll z).
\label{16}
\end{equation}
\\
While the ratio $\Pi _{X}/\rho _{X}$ is constant in this limit 
and may be of the order of unity, the
amount of $\Pi _{M}$ is much smaller than $\rho_{M}$.  For 
$\frac{\xi
}{3} + w _{X} <0$ the $X$ component (that is dynamically unimportant
for large $z$) looses energy which is transferred to the matter.
While $\Pi _{X}$ may be of the order of $\rho _{X}$, the quantity
$|\Pi _{M}|$ is negligible compared with $\rho_{M}$.  Since the
fractional quantities on the left-hand sides of Eqs. (\ref{15}) and
(\ref{16}) quantify the amount of the coupling, this means, the
dark matter does not feel the interaction, it is (almost) uncoupled.
As the evolution proceeds, $\Pi _{X}/\rho _{X}$ changes only slightly
to a present value
\\
\begin{equation}
-\frac{\Pi _{X}}{\rho _{X}} = \frac{r}
{1 + r }\left(\frac{\xi }{3} + w _{X}\right)
\qquad\qquad
\left(z =0\right)
 \ ,
\label{17}
\end{equation}
\\
whereas the corresponding ratio for the dark matter component becomes 
\\
\begin{equation}
\frac{\Pi _{M}}{\rho _{M}} = \frac{\frac{\xi }{3} + w _{X}}
{1 + r } 
\qquad\qquad\qquad\qquad
\left(z =0\right)
\ .
\label{18}
\end{equation}
\\
The point is that now the ratio $|\Pi _{M}|/\rho _{M}$ may also be of the order of 
unity, i.e., the dark matter fluid feels the coupling as well. {\it As far 
as the dark matter is concerned, the interaction has been switched on during the 
cosmic evolution.}  For $\xi=0$ we recover the relations of the 
previously discussed stationary solution \cite{interacting}. In the latter case 
the interaction does not depend on $z$.

\begin{figure}
\begin{center}
\includegraphics[width=0.7 \linewidth]{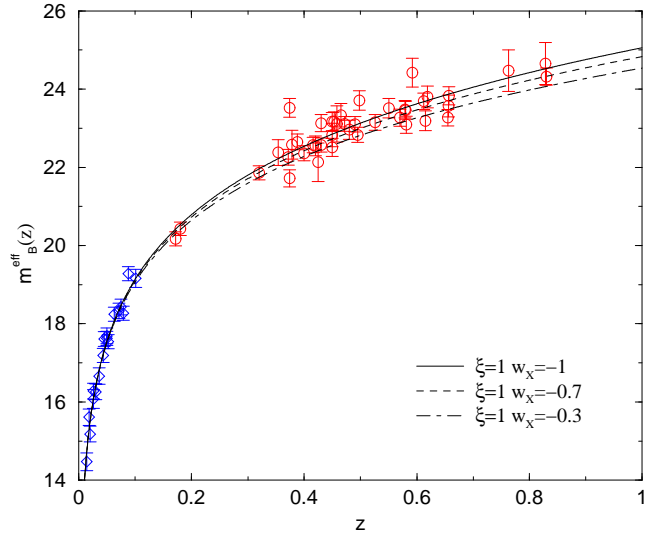}
\end{center}
\caption{Effective magnitudes for $r = \frac{3}{7} $ and $\xi = 1$}
\label{fig1}
\end{figure}

\begin{figure}
\begin{center}
\includegraphics[width=0.7\linewidth]{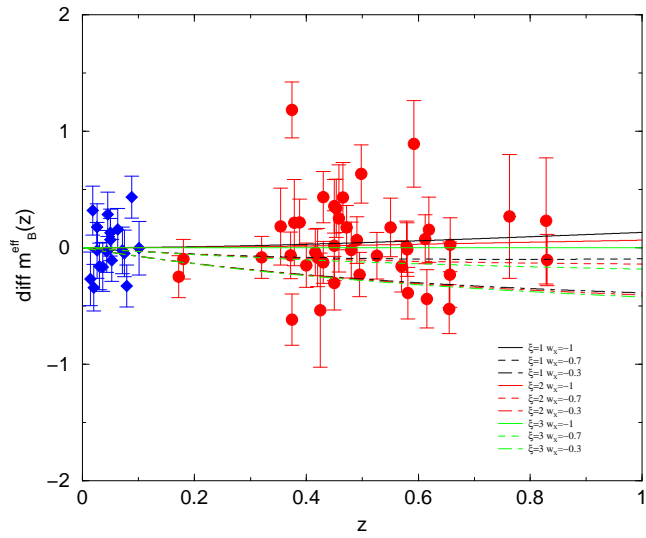}
\end{center}
\caption{Magnitude differences for various values of $\xi$ and $w_{X}$}
\label{fig3}
\end{figure}

Using the source terms corresponding to (\ref{6})  in the balances
(\ref{7}), the latter can be integrated. For the matter energy
density we find
\\
\begin{equation}
\rho _{M} = \rho _{M}\left(a _{0} \right)
\left[1+z \right]^{3 \left(1 + w _{X} \right)+\xi }
\left[\frac{1 + r \left(1+z \right) ^{\xi }}{1+r }\right]^{-1-\frac{3 w _{X}}{\xi }}\ .
\label{19}
\end{equation}
\\
The total energy density becomes 
\\
\begin{equation}
\rho  = \rho _{0}
\left[1+z \right]^{3 \left(1 + w _{X} \right)}
\left[\frac{1 + r \left(1+z \right) ^{\xi }}{1+r}\right]^{-\frac{
3 w _{X}}{\xi }}\ ,
\label{20}
\end{equation}
\\
where 
\begin{equation}
\rho _{0} = \frac{r+1}{r}\rho _{M}\left(a _{0} \right)\ .
\label{21}
\end{equation}
\\
Restricting ourselves to a universe with spatially flat sections, we obtain 
for the Hubble rate 
\\
\begin{equation}
H = \sqrt{\frac{8\pi G}{3}\rho _{0}}
\left(1+z\right)^{\frac{3}{2}\left(1+ w _{X} \right)}
\left[\frac{1+r \left(1+z \right)^{\xi }}{1+r}\right]
^{- \frac{3w _{X}}{2 \xi }} \ .
\label{22}
\end{equation}
\\
Likewise, the deceleration parameter  $q = - \ddot{a}/\left(a H ^{2} \right)$
can be expressed as
\\
\begin{equation}
q = \frac{1}{2}\frac{1+3w _{X} + r \left(1+z \right)^{\xi }}
{1 + r \left(1+z \right)^{\xi }}\ .
\label{23}
\end{equation}
\\
We find accelerated expansion for 
\\
\begin{equation}
1 + 3 w _{X} + r \left(z + 1\right)^{\xi } < 0 \ .
\label{24}
\end{equation}  
\\
The redshift $z _{acc}$ at which the acceleration starts, is determined by 
\\
\begin{equation}
1 + 3 w _{X} + r \left(z _{acc} + 1\right)^{\xi } =0 \ .
\label{25}
\end{equation}
\\
For $w _{X} = -1 $ (cosmological constant) we have accelerated expansion 
for
\\ 
\begin{equation}
\left(z + 1 \right)^{\xi } < \frac{2}{r}\ \quad\Rightarrow\quad
z _{acc} = \left(\frac{2}{r} \right)^{\frac{1}{\xi }} - 1 \ .
\label{26}
\end{equation}
\\
Both $H$ and $q$ depend on the set of parameters $r$, $w_{x}$, and $\xi$.
The luminosity distance
\\
\begin{equation}
d_{\rm L} = \left(1+z\right)\int_{0}^{z} 
\frac{\mbox{d}z}{H \left(z \right)}\ ,
\label{27}
\end{equation}
\\
as well as the angular distance $d_{\rm A}(z) = (1+z)^{-2} d_{\rm L}(z) $,
can be expressed in terms of them. The corresponding effective magnitude is 
\\
\[
m^{eff}_B = M_{Bf} + 5 \log (H_{0} d_{L}),
\]
\\
where we have chosen $M_{Bf} = -3.4$. Figure  \ref{fig1} 
depicts the effective magnitude $m^{eff}_{B}$ {\em vs} $z$ for $\xi = 1$ and 
different values of $w_{X}$ with $r = 3/7$.  
For $\xi = 3$ almost the same figure would appear as for $\xi = 1$. 
Figure \ref{fig3} shows the magnitude differences for various combinations of  
$\xi$ and $w_{X}$. It becomes apparent that a much richer set of data 
(hopefully to be provided by the SNAP satellite) will be needed to 
discriminate between these models. 

\section{Special case of a scaling cosmology}

The relations of the previous section considerably simplify for the special 
case $\xi=1$, $w_X = -1$. The total energy density (\ref{20}) reduces to
\\ 
\begin{equation}
\rho  = \frac{\rho _{0}}{\left(1+r \right)^3}
\left[1 + r\frac{a_0}{a}\right]^{3}\ .
\label{29}
\end{equation}
\\
For early times, $a \ll a_{0}$, the energy density $\rho$ redshifts as dust, while for
late times, $a_{0} \ll a$, it tends to a constant value. \\
Likewise, the expressions for the components are 
\\
\begin{equation}
\rho_M  = \frac{\rho _{M,0}}{\left(1+r \right)^2}\frac{a_0}{a}
\left[1 + r\frac{a_0}{a}\right]^{2}\ ,
\qquad  
\rho_X  = \frac{\rho _{M,0}}{r\left(1+r \right)^2}
\left[1 + r\frac{a_0}{a}\right]^{2}\ .
\label{30}
\end{equation}
\\
Furthermore, the ratios which quantify the interactions among dark matter 
and dark energy become 
\\
\begin{equation}
\frac{\Pi _{M}}{\rho _{M}} = - \frac{2}{3}
\frac{1}{1 + r \frac{a_0}{a}} \qquad \mbox{and} \qquad
\frac{\Pi _{X}}{\rho _{X}} =  \frac{2}{3}
\frac{r \frac{a_0}{a}}{1 + r \frac{a_0}{a}}
 \ .
\label{31}
\end{equation}
\\
The effective equation of state for the $X$ component is
\\
\begin{equation}
P_X = - \frac{1 + \frac{1}{3} r\frac{a_0}{a}}{1 + r \frac{a_0}{a}} \rho _{X}
 \ .
\label{32}
\end{equation}
\\
Assuming again $r=3/7$, this corresponds to a change from 
\\
\begin{equation}
P_X \approx - \frac{1}{3} \rho _{X} 
\qquad\qquad\qquad\qquad
\left(z\gg1\right)
\label{33}
\end{equation}
\\
at early times to 
\\
\begin{equation}
P_X \approx - \frac{4}{5} \rho _{X} 
\qquad\qquad\qquad\qquad
\left(z=0\right)
\label{34}
\end{equation}
\\
at the present epoch. The effective pressure $P_M$ of the matter component changes from a negligible value at $z \gg 1$ to the present value 
\\
\begin{equation}
P_M = \Pi_M \approx - \frac{7}{15} \rho _{M} 
\qquad\qquad\qquad\qquad
\left(z=0\right)\ .
\label{35}
\end{equation}
\\
The interaction has the effect that {\it both} components have a negative effective pressure. 
 
The expression (\ref{29}) for $\rho$ has to be contrasted with the energy density 
\\ 
\begin{equation}
\rho_{(\Lambda CDM) }  = \frac{\rho _{0}}{\left(1+r \right)}
\left[1 + r\left(\frac{a_0}{a}\right)^{3}\right]\ 
\label{36}
\end{equation}
\\
for the $\Lambda$CDM model. The sum of different powers in the latter is replaced 
by the power of a sum in our present model. The interaction in our model
makes $\rho_M$ decay at a lower rate than in the uncoupled case. 
The dark energy density $\rho_X$, on the other hand, which would remain 
constant without interaction, decays as well as a consequence of the transfer of 
energy to the matter component. This feature is familiar from decaying
cosmological constant models 
(see \cite{LiMa,GunzMaNe,vogue,ZB01}).

The solution of the Friedmann equation with $\rho$ from (\ref{29}) is
\\
\begin{eqnarray}
\frac{1}{2}
\left[\frac{1}{3}
\frac{\kappa\rho _{0}}{\left(1+r \right)^3}
\right]^{1/2} \left(t - t_0\right)
&=&  
\frac{1}{\sqrt{1+r}}
\left[1 - x
\sqrt{\frac{1+r}{1+ x}}\right] \nonumber\\
&& + \ln{\left\{\frac{\sqrt{r}}{1 + \sqrt{1+r}}
\left[x + \sqrt{x^2 + 1}\right] \right\}}
 \ ,
\label{37}
\end{eqnarray}
\\
where $x=\sqrt{\frac{a}{ra_0}}$.  In the limit $a\ll a_0$ we consistently 
recover the dust behavior $a\propto t^{2/3}$, while the scale factor 
approaches an exponential growth  for $a\gg a_0$. 
For the present special case $w_X =-1$, $\xi=1$ the integral in (\ref{27}) 
may be performed explicitly and  yields 
\\
\begin{equation}
d_{\rm L} = \frac{2}{H_0}\left(1+z\right)\frac{1+r}{r}
\left[1 - \frac{1}{\sqrt{1+ \frac{rz}{1+r}}}\right]  \ .
\label{38}
\end{equation}
\\
For small redshifts we obtain, up to third order in $z$, 
\\
\begin{equation}
H_0^{-1}d_{\rm L} \approx z 
\left[1 + \frac{1 + \frac{r}{4}}{1 + r}z 
- \frac{1}{8}\frac{r}{\left(1+r\right)^2}\left(6 + r\right)z^2\right]  \ .
\label{39}
\end{equation}
\\
Up to second order this expression coincides with the corresponding result for the 
$\Lambda$CDM model. Differences occur only in the $z^3$ term. Here, 
the factor $\left(6 + r\right)$ in the last expression replaces the factor 
$\left(10 + r\right)$ of the $\Lambda$CDM universe. 
The presently available SNIa data cannot discriminate between both models.  
Our model shares the merits of the $\Lambda$CDM model but 
at the same time alleviates the coincidence problem. 

\section{Conclusions}
Scaling solutions of the type $\rho_{M}/\rho_{X} = r (a_{0}/a)^{\xi}$
seem to be promising tools to deeper analyze the relationship between the 
two forms of energy dominating the current evolution of the universe,
namely, dark matter and dark energy. We showed that a suitably chosen 
interaction between them can lead to any scaling behavior of the 
mentioned form.  
In the specific case $w_{X} = -1, \xi = 1$ the
dynamics can be analytically integrated. 
Since the luminosity distance in this model differs from that of the 
$\Lambda$CDM model only in third order in the redshift parameter $z$, it 
fits the present observations as well as the $\Lambda$CDM model does. 
Further, the Einstein-de Sitter expansion law for a dust universe is 
recovered for large redshifts. 
On the other hand, the coincidence problem, although not solved, 
is less severe than for the $\Lambda$CDM universe, which can be traced back 
to a continuous transfer of energy from the $X$ component 
to the CDM fluid.    

\noindent
Wile the available observational data are insufficient to discriminate 
between the models, it is to be expected that the SNAP satellite will 
provide us with a wealth of high redshift supernovae data able to do 
the job. Likewise, complementary observations regarding the angular
distance between quasar pairs \cite{alcock} and the evolution of
cluster abundances \cite{abundances} will further constraint 
the set of parameters entering the scaling models.

\section*{Acknowledgements}
The authors are indebted to David Rapetti for helping us in
preparing the figures. This work was partially support by the NATO grant
PST. CLG.977973, and the Spanish Ministry of Science and Technology under
grant BFM 2000-C-03-01 and 2000-1322.

\end{document}